\documentclass[12pt]{iopart}

\usepackage{booktabs}
\usepackage{graphicx}
\usepackage{amstext}
\usepackage[backref=none]{hyperref} 
\usepackage[usenames,dvipsnames]{color}
\definecolor{MyDarkBlue}{rgb}{0,0.1,0.7}
\hypersetup{pdfborder={0 0 0},colorlinks,breaklinks=true,
  urlcolor={MyDarkBlue},citecolor={MyDarkBlue},linkcolor={MyDarkBlue}
}
\begin{document}

\title[PAOs: electronic accuracy and soft-mode distortions in ABO$_3$ perovskites]{The pseudoatomic orbital basis: electronic accuracy and soft-mode distortions in ABO$_3$ perovskites}

\author{Jack S. Baker$^{1, 2}$, Tsuyoshi Miyazaki$^3$ \& David R. Bowler$^{1, 2, 3}$}

\address{$^1$London Centre for Nanotechnology, UCL, 17-19 Gordon St, London WC1H 0AH, UK \\
         $^2$Department of Physics \& Astronomy, UCL, Gower St, London WC1E 6BT, UK \\
         $^3$International Centre for Materials Nanoarchitectonics (MANA), National Institute for Materials Science (NIMS), 1-1 Namiki, Tsukuba, Ibaraki 305-0044, Japan}
\ead{\textcolor{blue}{jack.baker.16@ucl.ac.uk}, \textcolor{blue}{MIYAZAKI.Tsuyoshi@nims.go.jp} \textcolor{blue}{drb@ucl.ac.uk}}
\vspace{10pt}
\begin{indented}
\item[]March 2020
\end{indented}

\begin{abstract}

The perovskite oxides are known to be susceptible to structural distortions over a long wavelength when compared to their parent cubic structures. From an \textit{ab initio} simulation perspective, this requires accurate calculations including many thousands of atoms; a task well beyond the remit of traditional plane wave-based density functional theory (DFT). We suggest that this void can be filled using the methodology implemented in the large-scale DFT code, \textsc{CONQUEST}, using a local pseudoatomic orbital (PAO) basis. Whilst this basis has been tested before for some structural and energetic properties, none have treated the most fundamental quantity to the theory, the charge density $n(\mathbf{r})$ itself. An accurate description of $n(\mathbf{r})$ is vital to the perovskite oxides due to the crucial role played by short-range restoring forces (characterised by bond covalency) and long range coulomb forces as suggested by the soft-mode theory of Cochran and Anderson. We find that modestly sized basis sets of PAOs can reproduce the plane-wave charge density to a total integrated error of better than $0.5\%$ and provide Bader partitioned ionic charges, volumes and average charge densities to similar degree of accuracy. Further, the multi-mode antiferroelectric distortion of PbZrO$_3$ and its associated energetics are reproduced by better than 99\% when compared to plane-waves. This work suggests that electronic structure calculations using efficient and compact basis sets of pseudoatomic orbitals can achieve the same accuracy as high cutoff energy plane-wave calculations. When paired with the \textsc{CONQUEST} code, calculations with high electronic and structural accuracy can now be performed on many thousands of atoms, even on systems as delicate as the perovskite oxides.

\end{abstract}

%
%
\submitto{Electron. Struct.}
%
\maketitle
%
%


\section{\label{sec:intro}Introduction}

The ABO$_3$ perovskite oxides are well known for their vast and rich variety of physical phenomena. These include interfacial two-dimensional electron gases \cite{Hwang2012, Huijben2009}, negative capacitance \cite{Zubko2016, MoraSer2006}, high-temperature superconductivity \cite{Maeno1994, Xiao1988} and many more. Many of these are linked to a plethora of responsible order parameters and their competition/coupling with one another \cite{Gareeva2009, Benedek2011}. Ferroelectric, ferromagnetic antiferroelectric \& antiferromagnetic order are all commonplace in the perovskite oxides as well as antiferrodistortions (rotation of the BO$_6$ octahedra) and Jahn-Teller distortions. Some of these features are also known to coexist with one another giving rise to the phenomena of multiferrocity \cite{Spaldin2019}. In simulation, the onset of many different competing order parameters can create a myriad of distinct local minima with similar energetics. It is then of paramount importance that our simulation methodology produces accurate results such that we can distinguish them from one another in their energetics but also accurately resolve their electronic \& structural properties. \par

In addition to the requirement of high accuracy, the perovskite oxides present structural and magnetic features over a long wavelength requiring first principles simulations of thousands of atoms. For example, thin ferroelectric films are known to form \textit{flux-closure domains} as a compensation mechanism for the depolarising field \cite{Streiffer2002, Fong2004}. The domain period in these films increases also with the depth of the film in question (the well known Kittel scaling law) thus requiring simulations in excess of a thousand atoms. In the case of multiferroic BiFeO$_3$, the competition between various exchange interactions manifests in the softening of a 64nm non-collinear spin-cycloid \cite{Sosnowska1982, Burns2019} and a unit cell of $\sim$1000 atoms. Current simulations often bypass this fact and approximate this complex magnetic order as simple G-type antiferromagnetism. Studies of solid-solution families (AB$_{x}$C$_{1-x}$O$_3$, $(1-x)$ABO$_3-x$CDO$_3$ and more) are popular in the field. Large supercell calculations can offer realistic experimental order of these alloys which can improve upon the accuracy of structural distortions found in smaller supercells and approximations like the virtual crystal approximation \cite{Baker2019, Bellaiche2000, Ramer2000}. Further, longer wavelength dynamical instabilities are found to be competitive in the important piezoelectric solid solution PbZr$_{0.5}$Ti$_{0.5}$O$_3$ (PZT 50/50) requiring a large number of atoms to simulate the energetics \cite{Baker2019}.

Electronic structure calculations based on density functional theory (DFT) employing the plane-wave pseudopotential method \cite{singh2006planewaves, Lejaeghere2016} are known to achieve accurate results. This is in part due to the systematic, variational nature of the the plane-wave basis where increasing the number of basis functions is \textit{guaranteed} to increase the level to which your calculations are converged. This method is not without its drawbacks. A plane-wave by itself (the solution of the free electron) bears little to no resemblance to the Kohn-Sham orbitals of the systems they are intended to represent. This is especially true for the localised 3d electrons of the transition metals, responsible for magnetic and orbital order. It is for this reason that many thousands of plane-waves are required in the basis set expansion at a great computational cost. Further, plane-waves span the whole of the simulation cell which introduces wasteful calculations on the grid for systems including a vacuum region. These issues can be bypassed by replacing plane-waves with physically intuitive local basis sets of pseudoatomic orbitals (PAOs) \cite{Kenny2000, soler2002siesta, Sankey1989, torralba2008pseudo, Louwerse2012}. These are atomic-like orbitals for which the radial part is solved in the pseudopotential of each ionic species \cite{Junquera2001, Anglada2002}. PAOs are now regularly used in the \texttt{Siesta} \cite{soler2002siesta}, \texttt{OpenMX} \cite{Ozaki2003} \& \textsc{CONQUEST} \cite{bowler2006recent} codes, the last of which is employed in this work. The construction and generation of such a basis is described in section \ref{PAOGen:level2}. \par

Our PAOs are designed with a cut-off in real space (where the basis function becomes zero) motivated by the desire to employ efficient sparse matrix algebra with high parallel efficiency \cite{Bowler2001}. Further, our formulation of DFT is based on the density matrix $\rho(\mathbf{r}, \mathbf{r^{\prime}})$. Should we choose to truncate the range of this matrix (a requirement should we wish to use the linear scaling mode of operation), we are physically supported by the principle of near-sightedness; the assertion that the density matrix $\rho(\mathbf{r}, \mathbf{r^{\prime}})$ decays to zero as $|\mathbf{r} - \mathbf{r^{\prime}}| \rightarrow \infty$ \cite{Kohn1996}. Complete with a change in algorithm (the scope of which is beyond this work but discussed in references \cite{bowler2012methods, bowler2006recent}), this allows the well known $\mathcal{O}(N^3)$ scaling wall (where $N$ is the number of atoms in the simulation) in standard DFT to be broken and replaced with a code which now scales as $\mathcal{O}(N)$. This method paves the way for full electronic structure calculations on systems of many thousands of atoms (or even millions \cite{Bowler2010}), well beyond what is possible with conventional plane-wave methods. The \textsc{CONQUEST} code has recently become publicly available with an MIT licence \cite{CQRelease2020}. \par

The accuracy of the PAO basis has been reported in previous works \cite{torralba2008pseudo, Louwerse2012, Bowler2019, Lee2007, Otsuka2008} including calculations of the structural parameters (bulk moduli and lattice constants) for this set of pseudopotentials \cite{Bowler2019}. Notably, none have reported on the effects to the most fundamental quantity in DFT; the charge density $n(\mathbf{r})$ itself. An accurate account of $n(\mathbf{r})$ is of a high importance for the perovskite oxides. Not only because applications require it (like the simulation of 2DEGs \cite{Yin2015}) but because of the possible ramifications for the soft-mode theory of ferroelectricity \cite{Cochran1959, Cochran1960}. That is, the  ferroelectric transition is governed by a zone centre dynamical instability driven by the competition of short range covalent forces (preferring cubic symmetry) and long range Coulomb forces (favouring the ferroelectric state). The charge density $n(\mathbf{r})$ (and its derived quantities) is clearly a probe of bond covalency \cite{Howard2003, Zhang2009} whilst electron-electron Coulomb terms feature explicit dependence on $n(\mathbf{r})$ in the calculation of the Hartree potential \cite{Martin2004}. \par

It is the purpose of this work to quantify the performance of PAOs versus the plane-wave pseudopotential method using calculations with the same pseudopotential. We compare the groundstate charge densities and the order parameters controlling ferroelectric \& antiferroelectric order. We do so by considering crystals of PbTiO$_3$ (PTO), PbZrO$_3$ (PZO) and two supercell arrangements of the solid solution PZT 50/50. PTO is a prototypical ferroelectric known to undergo a paraelectric to ferroelectric phase transition from cubic $Pm\bar{3}m$ to tetragonal $P4mm$ below 763K \cite{Nelmes1985}. This phase transition is known to come about from the softening of a zone centre lattice mode of irreducible representation (irrep) $\Gamma_4^-$ \cite{Nelmes1985}. In contrast, PZO undergoes a paraelectric to antiferroelectric phase transition from cubic $Pm\bar{3}m$ to orthorhombic $Pbam$ below 505K \cite{Tagantsev2013}. This transition is also thought to be caused by soft lattice modes \cite{Tagantsev2013, Mani2015, inguez2014, Fthenakis2017} but in this case has a more complex multi-mode description comprised primarily of R$_4^+$, $\Sigma_2$ \& S$_4$ modes \cite{Mani2015, inguez2014}. A small part of the distortion is also due to the softening of R$_5^+$, X$_3^-$ \& M$_5^-$ lattice modes \cite{Mani2015, inguez2014}. We quantify the amplitudes of each individual lattice mode (and strain modes, detailed in the supplemental material) in the phase transitions of both perovskites as well determining the associated energetics for each of the considered PAO basis sets. Since the energy differences associated with the perovskite oxides are generally small (a few meV/atom), this is a \textit{strict} test for the accuracy of PAOs.\par

The remainder of this work is now organized as follows. Within section \ref{CalcDetail:level2}, we outline the general simulation method for both the plane-wave and PAO DFT calculations and describe the phases of PTO, PZO \& PZT 50/50 considered in the study. In section \ref{PAOGen:level2} we describe the method for the generation of the PAO basis set and the details of the basis sets used in this work. Section \ref{ElecAcc:level2} provides charge density difference analysis between PAO calculations and plane-waves as well presenting the Bader analysis of the ionic charges, volumes and average densities. In section \ref{SoftModeAcc:level2} we compare the amplitudes of the soft-mode distortions responsible for the ferroelectric and antiferroelectric phase transitions in the PTO \& PZO including the energetics associated with crucial displacive modes. We also closely examine the energetics over the phase transition paths by steadily increasing mode amplitudes until a maximal value, then, cumulatively add the remaining important modes. We conclude this work in section \ref{Summary:level1} with a broad overview of our findings. This includes a discussion of the impact this work has on the topic of local basis sets and the promise of accurate and large-scale electronic structure calculations on the perovskite oxides. 

\section{\label{TheoMethod:level1}Theoretical method}
    \subsection{\label{CalcDetail:level2} Calculational details}
    
    Calculations are performed using two different implementations of DFT. Calculations using the plane-wave basis set are performed with the \texttt{ABINIT} code \cite{Gonze2016, Gonze2009} (\texttt{v8.10.2}) whilst calculations utilising PAOs are carried out using the \textsc{CONQUEST} code (\texttt{v1.0}) \cite{bowler2006recent, CQRelease2020} with the direct diagonalisation of the Hamiltonian matrix. Both codes are able to use the same norm-conserving pseudopotentials as produced by \texttt{ONCVPSP} \cite{hamann2013optimized} (\texttt{v3.3.1}) where input parameters were taken from the library (\texttt{v0.4}) on the \texttt{PseudoDojo} website \cite{van2018pseudodojo}. This library is known for its high accuracy results versus other pseudopotentials, projector augmented wave methods \& all-electron results as characterised in the well known DFT \textit{delta study} \cite{Lejaeghere2016}. These pseudopotentials are scalar-relativistic and include partial core corrections. The Pb 5d$^{10}$6s$^2$6p$^6$, Ti 3s$^2$3p$^6$4s$^2$3d$^{10}$, Zr 4s$^2$4p$^6$5s$^2$5d$^{10}$ and O 2s$^2$2p$^6$ orbitals are treated as valence in the pseudopotentials, but, in the \textsc{CONQUEST} calculations the $n=$3/4 states of Ti/Zr are treated as semi-core (see below). Exchange \& correlation is represented by the PBESol functional \cite{Perdew2008} as present in \texttt{Libxc} \cite{Marques2012} (\texttt{v3.0.0}). This functional is known to produce accurate structural properties for many crystals including the perovskite oxides \cite{Zhang2017}. For the \textsc{CONQUEST} calculations, we consider three different basis sets with increasing accuracy. These are the single-$\zeta$ plus polarisation (SZP), double-$\zeta$ plus double-polarisation (DZDP) and triple-$\zeta$ plus triple-polarisation (TZTP) basis sets respectively. Semi-core states (like those present in Ti/Zr) include only a single-$\zeta$ per $l$-channel. This approach allows us to study the effect of systematically adding an extra $\zeta$ per angular momentum channel. The basis sets and details of their generation are described fully in Section \ref{PAOGen:level2} \par
    
\begin{figure}
    \centering
       \includegraphics[width=\linewidth]{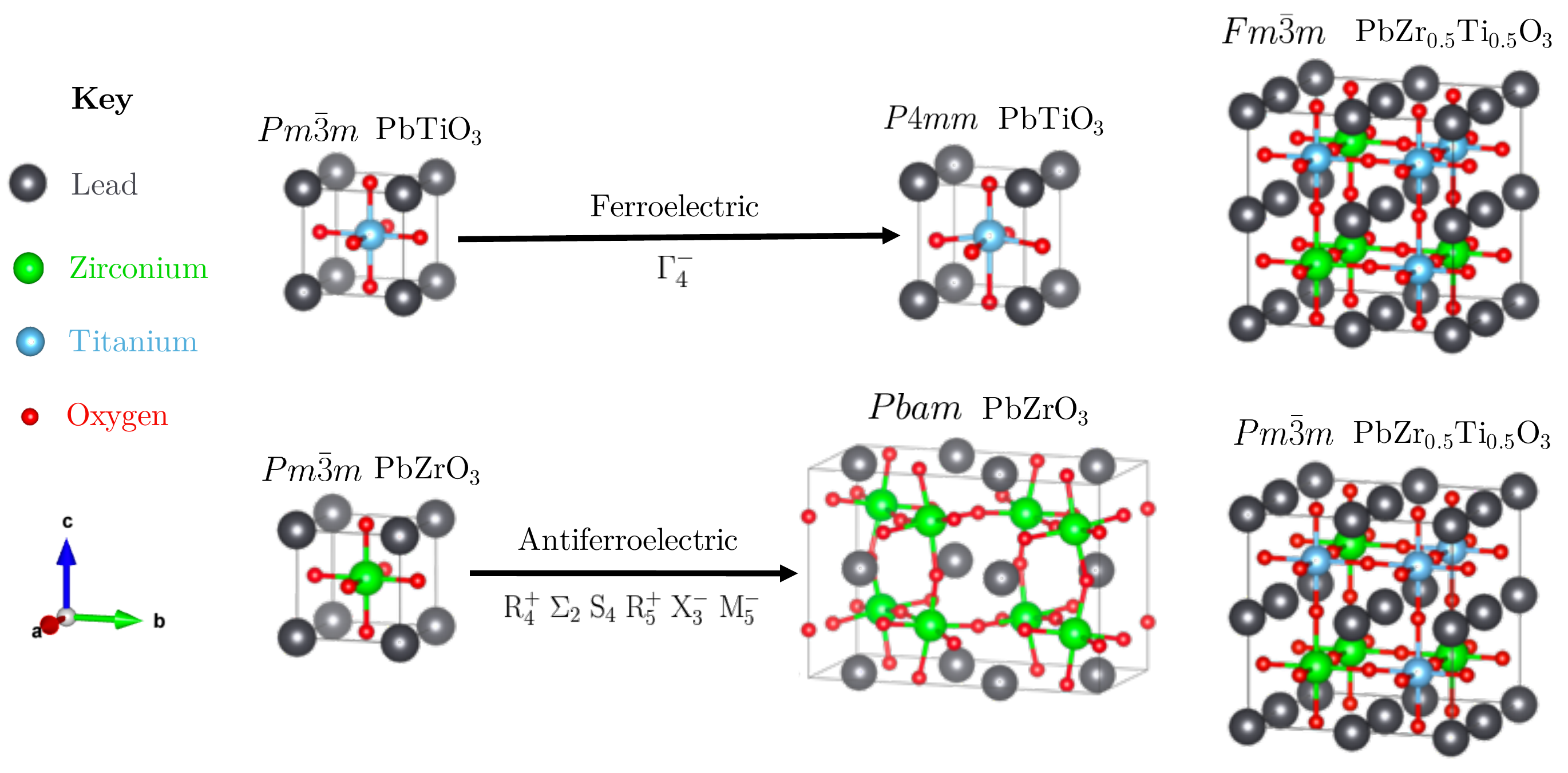}
        \caption{The crystal structures of PTO, PZO and PZT 50/50 considered in this work. Illustrated also are the paraelectric to ferroelectric and paraelectric to antiferroelectric phase transitions in PTO \& PZO, respectively. The responsible displacive soft lattice modes for the transitions are also indicated.}
        \label{fig:CrystalStructures}
    \end{figure}

    The different crystal structures treated in this study are shown in figure \ref{fig:CrystalStructures}. They display the cubic \& tetragonal phases of PTO, the cubic \& orthorhombic phases of PZO and two cubic arrangements of the PZT 50/50 solid solution. For cubic PZO, cubic PTO and tetragonal PTO, reciprocal space integrals are replaced with summations over a $9\times9\times9$ Monkhorst-pack \cite{monkhorst1976special} mesh whilst orthorhombic PZO and the cubic PZT 50/50 arrangements use a $7\times3\times5$ \& $5\times5\times5$ mesh respectively. For structural relaxation, \texttt{ABINIT} calculations use a 40Ha plane-wave cutoff and a 160Ha cutoff on the charge density grid whilst \textsc{CONQUEST} calculations use a 300Ha plane-wave-equivalent cutoff for both integrals on the grid and the charge density grid. These parameters were chosen that that total energies are converged to better than 1 meV/ABO$_3$ formula unit. The ionic positions are relaxed until the magnitude of the maximum force on all ions falls below $5\times 10^{-3}$ eV/\AA. Stresses are also relaxed until all elements of the Cartesian stress tensor fall below $1\times 10^{-4}$ GPa. This process is repeated for all considered crystals and for each basis set. \par
    
    For calculations assessing the charge density (section \ref{ElecAcc:level2}), we use a finer charge density grid with $100\times100\times100$ grid points/ABO$_3$ formula unit. For the orthorhombic $\sqrt{2} \times 2\sqrt{2} \times 2$ PZO unit cell, we use $150\times300\times200$ grid points. Each of these finer charge density calculations, for each basis, are performed using the optimised plane-wave structure of each crystal. In order to assign ionic charges, volumes and average ionic densities, we use the Bader partitioning scheme as implemented in the \texttt{bader} code \cite{Yu2011, Tang2009, Sanville2007, Henkelman2006} (\texttt{v1.03}). This code partitions individual atoms in crystals using the zero-flux surface of the charge density. This is a 2-D surface for which the charge density is at a minimum perpendicular to the surface. We note that whilst there is no unequivocal definition for the assignment of ionic charge, we choose the Bader definition since it derives \textit{only} from $n(\mathbf{r})$ thus introducing no new variables to our analysis. We define also a total integrated electronic error designed to quantify the level of disagreement in $n(\mathbf{r})$ for the plane-wave and PAO calculations. This is defined by the integral
    \begin{equation}
            N_{\text{error}}^e = \int |n_{\text{PAO}}(\mathbf{r}) - n_{\text{PW}}(\mathbf{r})|d\mathbf{r}
            \label{eq:NeError}
    \end{equation}
    for plane-wave/PAO electronic charge density $n_{\text{PW}}(\mathbf{r})$/$n_{\text{PAO}}(\mathbf{r})$.
    
    In section \ref{SoftModeAcc:level2} we assess the amplitudes of individual soft lattice modes in the phase transitions of PTO and PZO. To do so, we use the group symmetry analysis software made available in the \texttt{ISOTROPY} suite, in particular \texttt{ISODISTORT} \cite{Campbell2006} (\texttt{6.7.0}). This code is able to perform mode decompositional analysis when provided with the cubic $Pm\bar{3}m$ parent structures and the distorted daughter structures of PTO and PZO. In our calculations, the parent structures are the relaxed cubic $Pm\bar{3}m$ crystals for each basis set and daughter structures are the relaxed ferroelectric tetragonal PTO and antiferroelectric orthorhombic PZO cells for each basis.
    
    \subsection{\label{PAOGen:level2}Generation of pseudoatomic orbitals}
    
    PAOs are a local basis with a simple construction of a radial function $ R_{nl\zeta}(\mathbf{r})$  multiplied by an appropriate spherical harmonic $Y_m^l(\mathbf{\hat{r}})$.
        \begin{equation}
            \chi_{nlm\zeta}(\mathbf{r}) = R_{nl\zeta}(\mathbf{r})Y_m^l(\mathbf{\hat{r}})
        \end{equation}
    for principle quantum number $n$, orbital angular momentum $l$ and projection of orbital angular momentum $m$. The last subscript $\zeta$ is related to the number of functions per $l$-channel. Increasing the number of zetas adds flexibility to basis set and improves the accuracy in a non-variational manner. Since the spherical harmonics are analytic functions, the responsibility of the PAO generation code is to solve for the radial functions only. There is a question in this process with regards to how we should apply their confinement. The \texttt{Siesta} code introduced the concept of a uniform energy shift in the eigenvalues of the radial Schrödinger equation. This allows for a consistent definition of confinement for all angular momentum channels across all PAOs. Whilst our approach stems from this, we have implemented two variations for doing so within the \textsc{CONQUEST} PAO generator code  (\texttt{v1.02}). One approach is that of \textit{equal energies} where all functions with the same zeta share an energy shift. To increase flexibility of multiple zeta basis sets, one of the zetas should have a large confinement energy to create a highly confined function whilst others should have progressively less confinement. Another approach is \textit{equal radii}. Here, we solve all the radial functions at a given energy shift for given $\zeta$ then take the mean of the resulting radii. This radius is then used for the given $\zeta$. It was found in a recent study that whilst both methods showed good agreement with plane-wave calculations, the equal radii approach was found to give slightly better results for lattice constants and bulk moduli versus the equal energies method \cite{Bowler2019}.

    \begin{table}[]
        \centering
              \begin{tabular}{@{}cccc@{}}
              \toprule \toprule
               & \multicolumn{3}{c}{Cut-off radius, $r_c$ {[}$\text{\AA}${]}} \\ \midrule
               Species & $\zeta = 1$        & $\zeta = 2$        & $\zeta = 3$        \\ \midrule
               Pb      & 3.625              & 2.826              & 2.000              \\
               Zr      & 4.667              & 3.551              & 2.392              \\
               Ti      & 4.355              & 3.270              & 2.153              \\
               O       & 2.572              & 1.979              & 1.365              \\ \bottomrule \bottomrule
               \end{tabular}
        \caption{The cut-off radii $r_c$ for each $\zeta$ component of the radial functions which construct the basis sets used in this work by the \textit{equal radii} method. Note that the equal radii method produces radial functions whose cutoff does \textit{not} vary with $l$-channel. Semi-core states are not included here.}
        \label{tab:AvgRadii}
     \end{table}
    
    To generate the radial functions for the SZP, DZDP \& TZTP basis sets in this study, we use the \textsc{CONQUEST} PAO generator (\texttt{v1.02}) operating under the equal radii scheme. Polarisation functions are solved in a peturbative manner by considering the effect of a finite local electric field acting on the highest valence state \cite{Artacho1999}. We use the default setting, applying confinement energies of 2 eV, 0.2 eV and 0.02 eV for the first, second and third zetas respectively. The average radii for each zeta for each species for the valence states are shown in table \ref{tab:AvgRadii}. The semi-core Zr and Ti radial functions are highly confined with the Zr 4s \& 4p radii being 1.953 $\text{\AA}$ \& 2.281 $\text{\AA}$ respectively. Ti 3s \& 3p semi-core states are cut off at 1.757 $\text{\AA}$ \& 2.001 $\text{\AA}$ (The radial functions are displayed in section 1 of the supplemental material). The equal radii method produces slightly more compressed functions than the equal energies method which are generally more efficient (due to their smaller cut-off radius) but may require integration on a finer grid due to their more exaggerated gradients. We emphasize once more that the radial functions used in this work are the \textit{defaults} of the PAO generator code. Any results here then should be regarded as \textit{out-of-the-box} performance because in principle, it is possible to fit/optimize these functions for specific situations. For example, the approach made by the \texttt{Siesta} code is a downhill simplex minimisation of the total energy carried out on the material system to be studied (usually a solid-state or molecular system) with respect to the parameters of the PAO generation mechanism \cite{Junquera2001}. Another approach optimises the binding energy curve of dimers \cite{Oroya2020}. Whilst these methods can produce good results, the possibility of many local minima in the optimisation is an issue as is the overfitting of smaller PAO bases such that their transferability is diminished. In this respect, a large basis set of default (and more general) PAOs may be more transferable.
    
\section{\label{Results:level1} Results}
    \subsection{\label{ElecAcc:level2} Electronic accuracy}
    
\begin{figure}
    \centering
       \includegraphics[width=\linewidth]{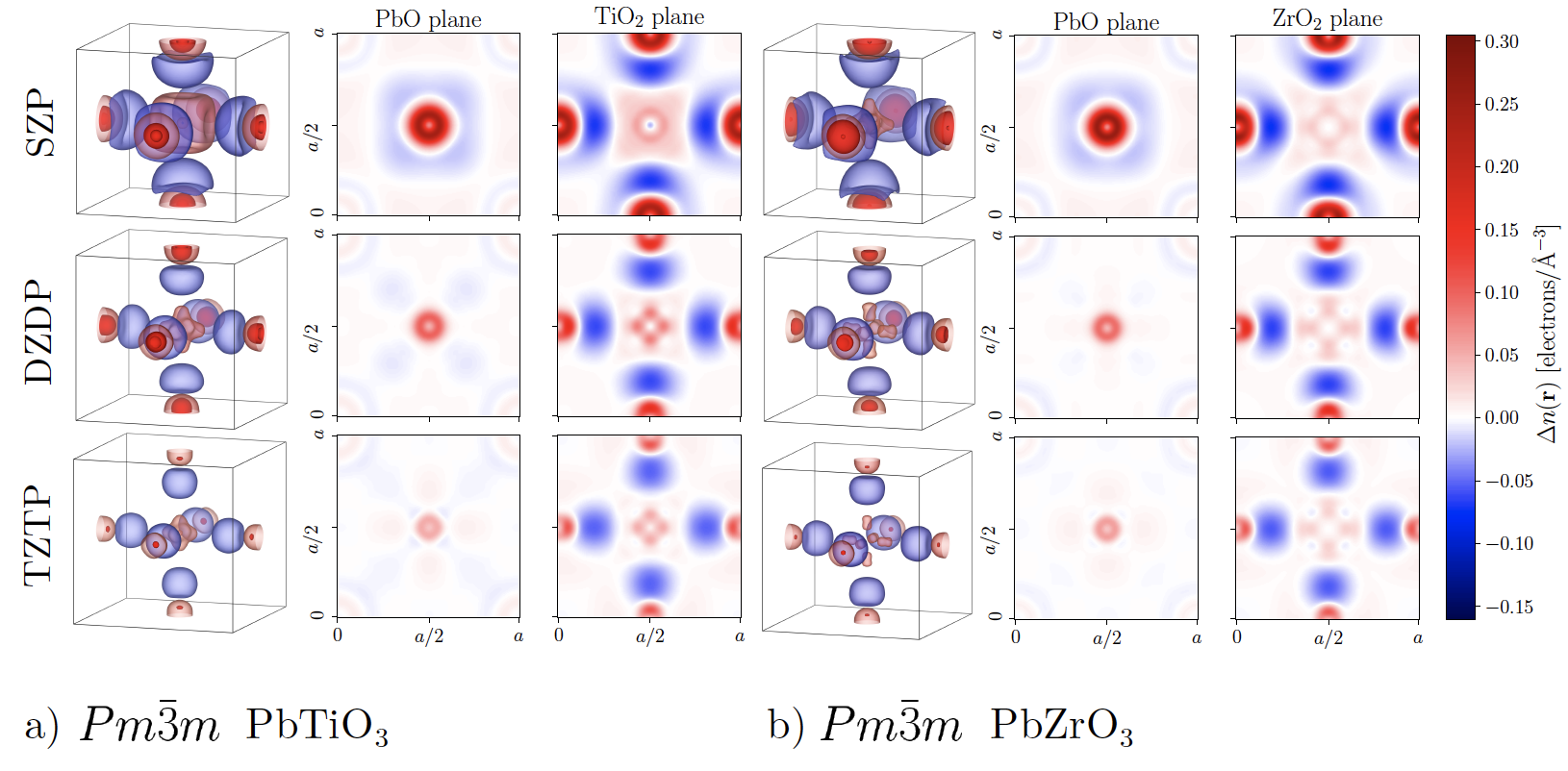}
        \caption{The charge density difference between PAO calculations and plane-waves for increasing PAO basis set size for cubic PTO (a) and cubic PZO (b). For each case, we display the full 3d isosurfaces and slices through the PbO and BO$_2$ planes. Isosurfaces are plotted at the +0.10 (dark red), +0.020 (light red) and -0.020 (blue) electrons/\AA$^{-3}$ levels.}
        \label{fig:PTOPZODiff}
    \end{figure}

\begin{figure}
    \centering
       \includegraphics[width=0.9\linewidth]{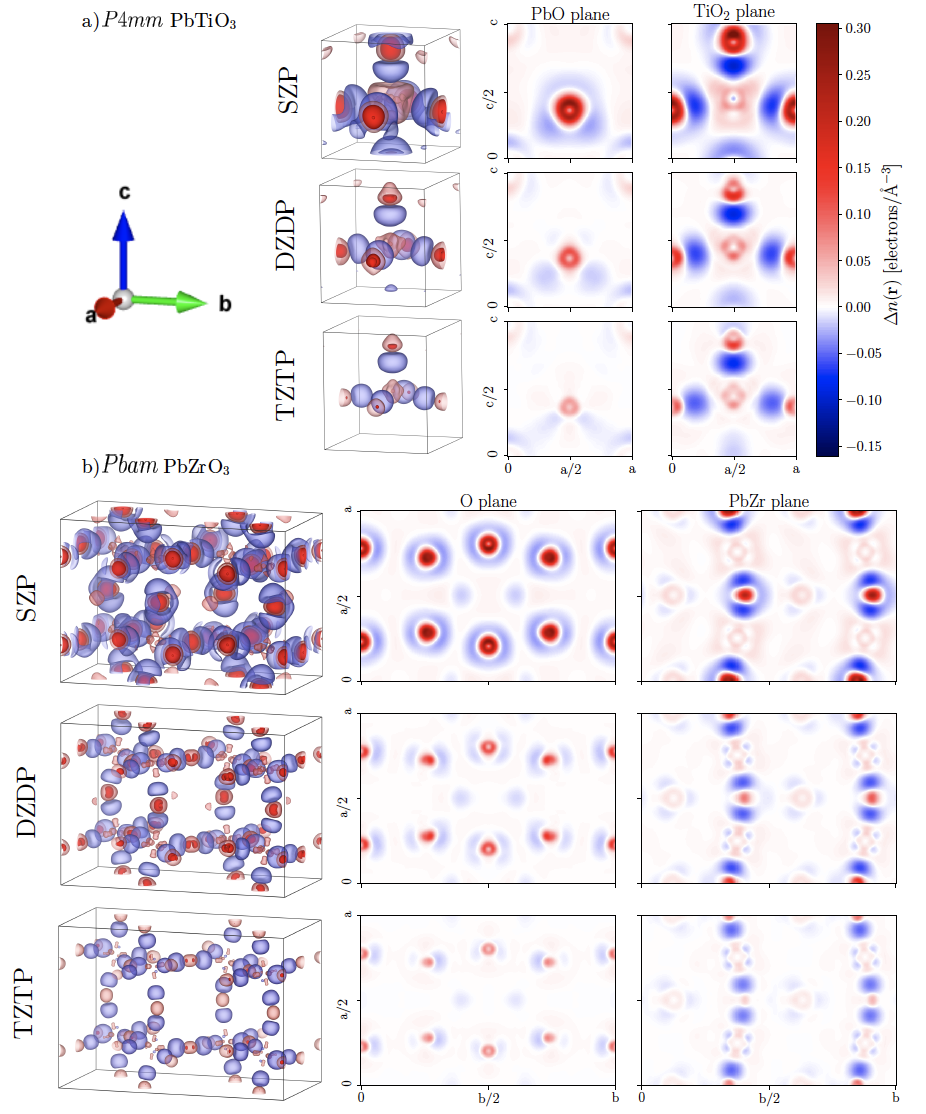}
        \caption{The charge density difference between PAO calculations and plane-waves for increasing PAO basis set size for tetragonal PTO (a) and orthorhombic PZO (b). For each case, we display the full 3d isosurfaces and selected  slices. Isosurfaces are plotted at the +0.10 (dark red), +0.020 (light red) and -0.020 (blue) electrons/\AA$^{-3}$ levels.}
        \label{fig:pbam_p4mmPTOPZODiff}
    \end{figure}

\begin{figure}
    \centering
       \includegraphics[width=0.75\linewidth]{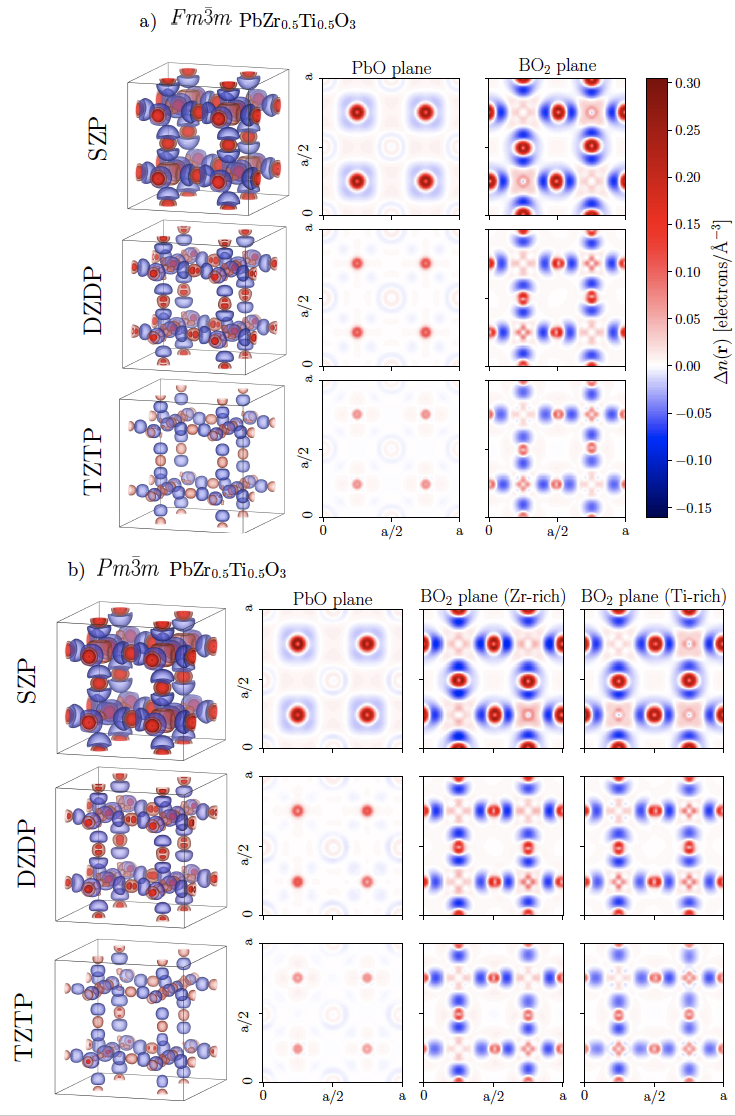}
        \caption{The charge density difference between PAO calculations and plane-waves for increasing PAO basis set size for $Fm\bar{3}m$ cubic PZT 50/50 (a) and $Pm\bar{3}m$ cubic PZT 50/50 (b). For each case, we display the full 3d isosurfaces and selected  slices. Isosurfaces are plotted at the +0.10 (dark red), +0.020 (light red) and -0.020 (blue) electrons/\AA$^{-3}$ levels. }
        \label{fig:PZTChdenDiff}
    \end{figure}

    Figures \ref{fig:PTOPZODiff}, \ref{fig:pbam_p4mmPTOPZODiff} \& \ref{fig:PZTChdenDiff} show the charge density differences $\Delta n(\mathbf{r}) = n_{\text{PAO}}(\mathbf{r}) - n_{\text{PW}}(\mathbf{r})$ for each PAO basis set and each crystal structure shown in figure \ref{fig:CrystalStructures}. We show this quantity using both coloured isosurfaces and slices through chosen planes which are described in each figure. Before discussing the details of each case, we discuss some striking features shared by all cases. The range of $\Delta n(\mathbf{r})$ is similar between all crystals, extremal at around 0.25 electrons$/\text{\AA}^3$ with a very narrow region of negative $\Delta n(\mathbf{r})$, minimal at $ \approx 0.07$ electrons$/\text{\AA}^3$. Even when considering these extrema, their magnitudes are $\approx50\times$ smaller than the extrema of any given $n(\mathbf{r})$. This is even more apparent when considering the mean absolute of $\Delta n(\mathbf{r})$ which is $\approx 0.01$ electrons$/\text{\AA}^3$ for SZP falling to $\approx0.004$ electrons$/\text{\AA}^3$ by TZTP. This shows that even at first glance for all bases the electronic error versus plane-waves is small. It is also clear that these regions of maximal $\Delta n(\mathbf{r})$ are small in volume and isolated close to each ionic site, especially the O anions. Further, error is almost vanishing proximal to the Pb cations and Pb-O bonds suggesting this chemistry is well described by the default PAOs. Other than the aforementioned sites, $\Delta n(\mathbf{r}) \approx 0$ for the vast majority of the simulation box. Since all calculations are normalised to the same number of electrons (44/ABO$_3$ unit) the localised surplus of electrons close to ionic sites \textit{has to} result in an electron deficiency elsewhere. This manifests itself in two areas. Firstly, a small negative $\Delta n(\mathbf{r})$ appears in bonding areas, especially those characterising the BO$_6$ octahedra. Secondly, the remaining $\Delta n(\mathbf{r})$ spreads itself out as an even smaller negative background over the rest of the simulation box. Finally, we see that the effect of increasing the size of the PAO basis from SZP to DZDP results in a large reduction in $\Delta n(\mathbf{r})$. This effect is most clear when we examine the shrinking volumes enclosed by the isosurfaces present on any one of figures \ref{fig:PTOPZODiff}, \ref{fig:pbam_p4mmPTOPZODiff} or \ref{fig:PZTChdenDiff}. This effect can also be seen as we increase the basis set size from DZDP to TZTP but is less drastic. \par
    
    \begin{table}[]
\centering
\begin{tabular}{@{}ccccccc@{}}
\toprule \toprule
     & \multicolumn{6}{c}{$N_{\text{error}}^e$/ABO$_3$ unit (44 electrons) {[}e{]}}                                                   \\ \midrule
     & $Pm\bar{3}m$ & $Pm\bar{3}m$ & $P4mm$    & $Pbam$    & $Fm\bar{3}m$                & $Pm\bar{3}m$                \\
     & PTO    & PZO    & PTO & PZO & PZT 50/50 & PZT 50/50 \\ \midrule
SZP  & 0.719        & 0.768        & 0.730     & 0.772     & 0.739                       & 0.743                       \\
DZDP & 0.358        & 0.382        & 0.380     & 0.397     & 0.369                       & 0.371                       \\
TZTP & 0.246        & 0.276        & 0.257     & 0.282     & 0.257                       & 0.259                       \\ \bottomrule \bottomrule
\end{tabular}

\caption{The total integated electronic error $N_{\text{error}}^e$ (normalized per 5-atom ABO$_3$ perovskite unit, containing 44 electrons) as defined in eq \ref{eq:NeError} for the SZP, DZDP \& TZTP PAO basis sets for each of the considered structures in figure \ref{fig:CrystalStructures}.}
\label{tab:NeError}
\end{table}
    
    Table \ref{tab:NeError} shows the total integrated electronic error as defined in eq \ref{eq:NeError}. Much like the range of $\Delta n(\mathbf{r})$, the magnitude is similar across all crystals and decreases as we increase the basis set size. This improvement is once again greater between the SZP \& DZDP basis sets as compared to the drop in $N_{\text{error}}^e$ between DZDP \& TZTP.  What is notable, however, is that there is a noticeable (but small) gain in $N_{\text{error}}^e$ as we break $Pm\bar{3}m$ symmetry for cubic PTO \& PZO to the distorted $P4mm$ \& $Pbam$ phases respectively. This gain is comparable for both compounds. This effect can be explained by a slight rigidity for each PAO local to sites which become low symmetry. The basis must now adapt to the now distorted environment in a more asymmetrical manner which doesn't perform as well as the same process at a higher symmetry site. This effect can be seen clearly in figure \ref{fig:pbam_p4mmPTOPZODiff}a when examining the TiO$_2$ panel for the SZP basis set. Close to the O 1b Wyckoff site we see firstly that $\Delta n(\mathbf{r})$ is now asymmetrical in the plane perpendicular to the the pseudocubic c-axis as compared to the same panel in figure \ref{fig:PTOPZODiff}a where $\Delta n(\mathbf{r})$ is symmetrical. We see also the $\Delta n(\mathbf{r})$ is now greater (and extremal) in the upper region of the O 1b site which is a primary source of the increase in the total integrated error as we break cubic symmetry. We see that for the two PZT 50/50 configurations that $N_{\text{error}}^e$ is comparable since both arrangements have a similar, cubic symmetry. It is also interesting to point out that the mean of $N_{\text{error}}^e$ between cubic PTO \& PZO for any basis \textit{very closely} mirrors the value of $N_{\text{error}}^e$ for either of the PZT 50/50 configurations. This suggests that the electronic structure local to PTO \& PZO units in the alloy is similar to that of the pure compound. This is supported further when examining the BO$_2$ panels of figure \ref{fig:PZTChdenDiff} where local PTO \& PZO units are easily discernible when compared with the BO$_2$ panels of figure \ref{fig:PTOPZODiff}. This suggests that approximations (like the virtual crystal approximation) designed to circumvent the need for large supercell calculations of alloys are unable to accurately account for local electronic structure. \par
    
\begin{figure}
    \centering
       \includegraphics[width=\linewidth]{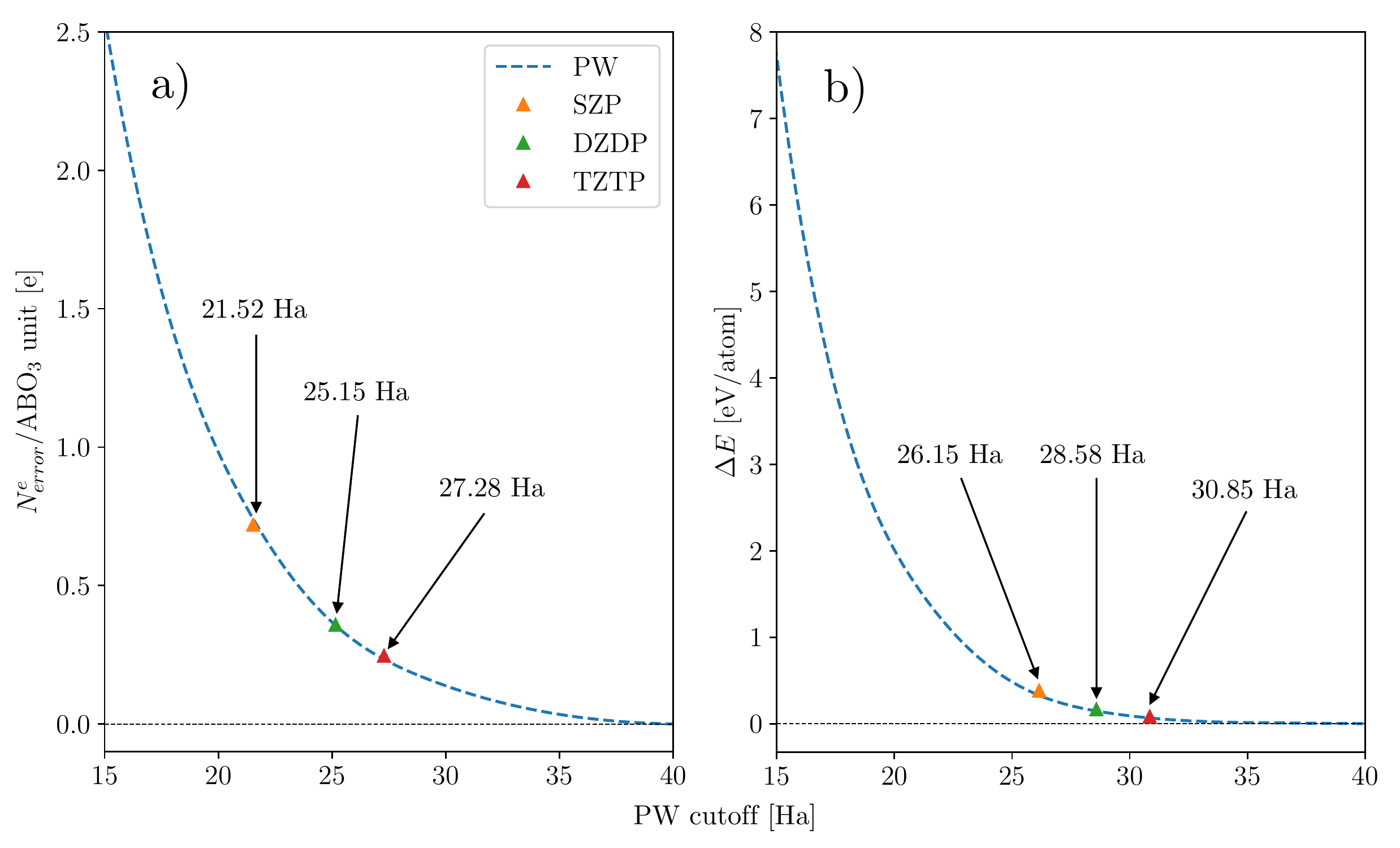}
       \caption{The convergence properties of plane-wave calculations where PAO calculations featuring \textit{the same error} have been overlaid for comparison. Calculations were performed on the $Pm\bar{3}m$ PTO structure. a) Convergence with respect to the total electronic error integral of eq \ref{eq:NeError} b) Convergence with respect to $\Delta E$, the energy difference between a given calculation and the energy obtained from the 40Ha plane-wave cutoff.}
       \label{fig:PAOVersusPW_ecut}
    \end{figure}

    Figure \ref{fig:PAOVersusPW_ecut}a quantifies $N_{\text{error}}^e$ for $Pm\bar{3}m$ PTO in terms of the same error produced by a given plane-wave cutoff energy using the plane-wave basis. We see that when increasing the number of PAOs to TZTP, we achieve the same accuracy as a 27.28 Ha cutoff plane-wave calculation. Figure \ref{fig:PAOVersusPW_ecut}b makes the same comparison but for the convergence of the total energy difference $\Delta E$. All PAO basis sets perform better using this metric (in particular, SZP makes a gain of +4.63 Ha in plane-wave cutoff energy) with TZTP achieving the same convergence in energy as a 30.85 Ha plane-wave cutoff. We note that these values are close to \textit{double} those reported by the PAO basis sets in the \texttt{Siesta} code \cite{soler2002siesta} but accept that some of this difference could be accounted for due to the softer (and lower accuracy) Troullier-Martins \cite{Troullier1991} type pseudopotentials used by the code and the difference in material system used for the study (bulk Si).  \par

\begin{table}[]
\centering
\scalebox{0.84}{\begin{tabular}{@{}ccccccccccccc@{}} \toprule \toprule

                                 & \multicolumn{4}{c}{$q_{\text{B}}$ {[}e{]}}      & \multicolumn{4}{c}{$V_{\text{B}}$ {[}$\text{\AA}^3${]}} & \multicolumn{4}{c}{$\bar{n}_{\text{B}}$ {[}$e/\text{\AA}^3${]}} \\ \midrule
                                 & \multicolumn{12}{c}{$Pm\bar{3}m$ PTO}                                                                                                                                 \\ \midrule
\multicolumn{1}{c|}{}            & SZP   & DZDP  & TZTP  & \multicolumn{1}{c|}{PW} & SZP     & DZDP    & TZTP    & \multicolumn{1}{c|}{PW}   & SZP       & DZDP      & TZTP       & \multicolumn{1}{c|}{PW}    \\ \midrule
Pb                               & 1.29  & 1.37  & 1.39  & 1.38                    & 18.81   & 18.40   & 18.39   & 18.41                     & 0.676     & 0.686     & 0.686      & 0.686                      \\
Ti                               & 2.01  & 2.08  & 2.12  & 2.15                    & 7.68    & 7.54    & 7.42    & 7.35                      & 1.300     & 1.315     & 1.331      & 1.344                      \\
O                                & -1.10 & -1.15 & -1.17 & -1.18                   & 11.22   & 11.40   & 11.45   & 11.46                     & 0.633     & 0.627     & 0.6326     & 0.626                      \\ \midrule
                                 & \multicolumn{12}{c}{$Pm\bar{3}m$ PZO}                                                                                                                                 \\ \midrule
Pb                               & 1.21  & 1.30  & 1.39  & 1.35                    & 21.89   & 21.42   & 20.83   & 21.06                     & 0.585     & 0.593     & 0.606      & 0.600                      \\
Zr                               & 2.31  & 2.39  & 2.43  & 2.47                    & 11.00   & 10.65   & 10.39   & 10.23                     & 0.881     & 0.903     & 0.921      & 0.931                      \\
O                                & -1.17 & -1.23 & -1.27 & -1.27                   & 12.70   & 12.97   & 13.25   & 13.23                     & 0.565     & 0.557     & 0.549      & 0.550                      \\ \midrule
                                 & \multicolumn{12}{c}{$P4mm$ PTO}                                                                                                                                       \\ \midrule
Pb                               & 1.27  & 1.35  & 1.39  & 1.37                    & 19.70   & 19.26   & 19.07   & 19.16                     & 0.646     & 0.657     & 0.661      & 0.659                      \\
Ti                               & 2.03  & 2.10  & 2.13  & 2.16                    & 7.83    & 7.67    & 7.56    & 7.48                      & 1.273     & 1.291     & 1.306      & 1.315                      \\
O                                & -1.10 & -1.15 & -1.17 & -1.18                   & 11.76   & 11.96   & 12.07   & 12.06                     & 0.603     & 0.597     & 0.594      & 0.595                      \\ \midrule
                                 & \multicolumn{12}{c}{$Pbam$ PZO}                                                                                                                                       \\ \midrule
Pb                               & 1.20  & 1.29  & 1.34  & 1.31                    & 21.24   & 20.70   & 20.52   & 20.66                     & 0.603     & 0.614     & 0.617      & 0.614                      \\
Zr                               & 2.37  & 2.47  & 2.51  & 2.55                    & 10.99   & 10.61   & 10.39   & 10.28                     & 0.876     & 0.898     & 0.913      & 0.920                      \\
O                                & -1.19 & -1.25 & -1.28 & -1.29                   & 12.79   & 13.10   & 13.23   & 13.22                     & 0.562     & 0.554     & 0.550      & 0.551                      \\ \midrule
                                 & \multicolumn{12}{c}{$Fm\bar{3}m$ PZT 50/50}                                                                                                               \\ \midrule
Pb                               & 1.24  & 1.34  & 1.39  & 1.35                    & 20.29   & 19.78   & 19.52   & 19.76                     & 0.629     & 0.640     & 0.646      & 0.640                      \\
Zr                               & 2.26  & 2.36  & 2.42  & 2.55                    & 10.91   & 10.54   & 10.28   & 9.96                      & 0.892     & 0.915     & 0.933      & 0.949                      \\
Ti                               & 2.04  & 2.12  & 2.15  & 2.18                    & 7.70    & 7.53    & 7.43    & 7.34                      & 1.294     & 1.313     & 1.325      & 1.338                      \\
O                                & -1.13 & -1.19 & -1.22 & -1.24                   & 11.87   & 12.13   & 12.28   & 12.27                     & 0.601     & 0.593     & 0.588      & 0.590                      \\ \midrule
                                 & \multicolumn{12}{c}{$Pm\bar{3}m$ PZT 50/50}                                                                                                               \\ \midrule
Pb                               & 1.24  & 1.34  & 1.39  & 1.36                    & 20.45   & 19.86   & 19.60   & 19.79                     & 0.624     & 0.637     & 0.643      & 0.639                      \\
Zr                               & 2.33  & 2.39  & 2.43  & 2.47                    & 10.64   & 10.43   & 10.18   & 10.03                     & 0.909     & 0.921     & 0.939      & 0.950                      \\
Ti                               & 2.04  & 2.11  & 2.15  & 2.18                    & 7.73    & 7.54    & 7.42    & 7.34                      & 1.289     & 1.311     & 1.327      & 1.337                      \\
O                                & -1.14 & -1.20 & -1.23 & -1.23                   & 12.01   & 12.27   & 12.42   & 12.39                     & 0.595     & 0.587     & 0.582      & 0.583                      \\ \midrule
$\epsilon$ {[}\%{]}              & -7.55 & -2.83 & -0.26 & -                       & 2.45    & 1.20    & 0.37    & -                         & -1.73     & -0.89     & -0.26      & -                          \\
$\epsilon^{\text{abs}}$ {[}\%{]} & 7.55  & 2.83  & 1.45  & -                       & 4.27    & 1.84    & 4.27    & -                         & 2.82      & 1.25      & 0.60       & -                          \\ \bottomrule \bottomrule
\end{tabular}}
\caption{The Bader partitioned ionic charges $q_B$, volumes $V_B$ and average valence charge densities $\bar{n}_B$ for each ionic species and each of the considered crystal structures calculated for each basis set. Where ionic species have symmetry in-equivalent sites, we take the mean value over all sites. We give also the mean relative error (MRE) $\epsilon$ and mean absolute relative error (MARE) $\epsilon^{\text{abs}}$ measured in \% error versus plane-waves.}
\label{tab:bader}
\end{table}

Table \ref{tab:bader} shows the quantities derived from Bader partitioning of the charge densities calculated using the plane-wave optimised geometries. These results reveal some fine features in the electronic structure. We see that for all basis sets, taking cubic PTO as an example, the Bader ionic charges $q_B$ (obtained from the difference in the pseudopotential valence charge with the Bader partitioned valence charge) \textit{do not} coincide well with the nominal charges (Pb$^{2+}$, Ti$^{4+}$, O$^{2-}$) which are used frequently in the literature as a convenience rather than an \textit{ab initio} assignment. We retrieve around half the nominal values suggesting a significant covalent character to the bonding, especially for the BO$_6$ octahedra. This is in contrast to the well known of phenomenon anomalous dynamical or \textit{Born effective} charges in the perovskite oxides, known to approximately \textit{double} the nominal charges of ions in the perovskite oxides \cite{Zhong1994}.\par

We see that there is a cation $\rightarrow$ anion negative charge transfer with increasing basis set size suggesting an increasing level of ionicity in bonding bringing $q_B$ closer to their nominal values. We see that the prediction of $q_B$ is rather underestimated for cations and overestimated for anions in the SZP basis set. This suggests that the electronegativity of O is underestimated and/or electrons on the metal ion sites are over-localised compared to plane-waves. This fact first appears to be at odds with the observed effect of an electron surplus near O anions in seen in figures \ref{fig:PTOPZODiff}, \ref{fig:pbam_p4mmPTOPZODiff} \& \ref{fig:PZTChdenDiff}. This is however rationalised as we also observe a decrease in the Bader volume $V_B$ for O anions which, in turn, increases the average valence charge density $\bar{n}_B$ (as seen in table \ref{tab:bader}) recovering the effect observed in the electron density difference plots. Whilst it can be seen that the Bader derived quantities are rather approximate (most notably in $q_B$) for the SZP basis, by DZDP (and \textit{certainly} by TZTP) they are in good agreement with the values obtained from plane-waves. This once again emphasizes the electronic accuracy achievable with the default PAOs. \par

We have also examined the possibility that the errors in $q_B$ could be an artefact of pressure using the particular case of $Pm\bar{3}m$ PZO. We can see from table \ref{tab:latticeconstants} that the optimised lattice constants for the PAO bases overestimate the plane-wave result by 0.99\%, 0.46\% and 0.34\% for the SZP, DZDP and TZTP basis sets respectively. If we then perform simulations at the plane-wave lattice constant (as was done for the results in table \ref{tab:bader}) this imposes an isotropic pressure of -5.28 GPa (SZP), -2.63 GPa (DZDP) and -1.46 GPa (TZTP). Because of this fact, we calculated the Bader quantities once more using the zero pressure lattice constants. Remarkably, $q_B$ changes only marginally (no more than $\pm 0.01e$). Since we are working at a larger volume, the $V_B$ must of course increase, but, the ratios of the cationic to anionic volumes remain constant. Since roughly the same amount of charge is now enclosed within a larger $V_B$, we naturally see a decreased $\bar{n}_B$ for all sites.

We see that the ratio of the cation to anion volumes is a decreasing function of basis set completeness, decreasing by $\approx 0.1$ for $V_B^B/V_B^O$ and $\approx 0.2$ for $V_B^{Pb}/V_B^O$ from SZP to TZTP. This implies that in the smaller basis set, O occupies a smaller ionic volume in comparison to the Pb and B-sites. This could result in small differences in lattice dynamics between the basis sets and could effect the Goldschmidt tolerance factor, depending explicitly on ionic radii \cite{Goldschmidt1926, Bartel2019}.

\subsection{\label{SoftModeAcc:level2} Soft-mode distortions}

\begin{figure}
    \centering
       \includegraphics[width=\linewidth]{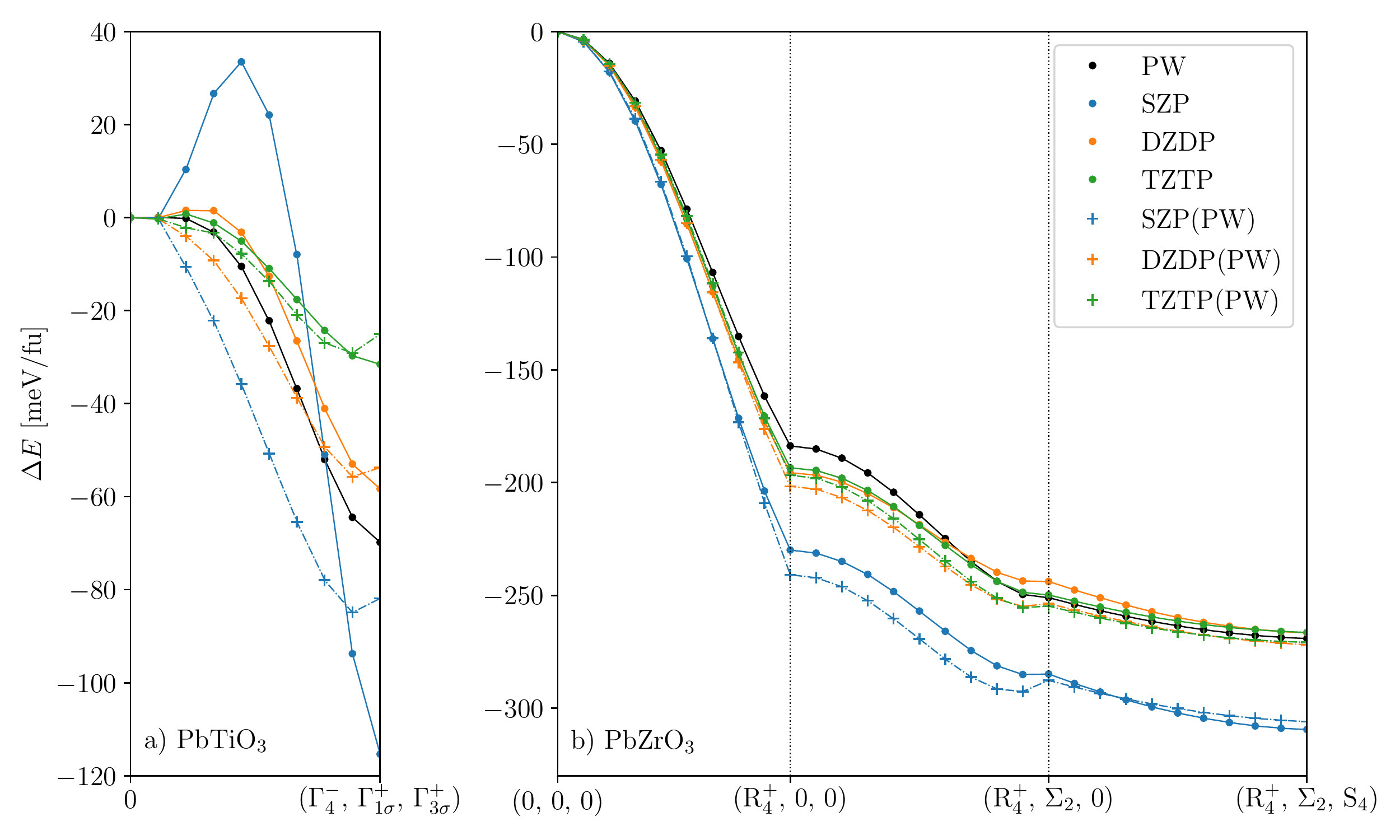}
        \caption{The phase transition energetics for PTO and PZO as function of normalised mode amplitude. Points/crosses are calculated data whereas lines are linear
        interpolations. Curves labelled with (PW) were calculated along the plane-wave optimised trajectory. a) Seven curves indicating phase transition paths for the zone centre distortions in PTO. Modes with the subscript $\sigma$ are strain modes coupled to the displacive $\Gamma_4^-$ mode. b) Seven curves describing the phase transition path for the three most important modes in the $Pbam$ PbZrO$_3$ phase transition. Each path begins with the optimised orthorhombic cell for each basis such that strain modes are "pre-frozen".}
        \label{fig:ModeEnergy}
    \end{figure}

In this section we consider the soft-modes known to drive the phase transitions in PTO and PZO. We consider the amplitude of each identifiable irrep in the relaxed structures for each basis set. We also consider the degree of energy lowering associated with each of these irreps and define phase transition energies. We display the displacive modes in Tables \ref{tab:PTODistortTab} and \ref{tab:PZODistortTab}. Strain modes influence the phase transition in PZO only by a small amount so we include only discussion of strain modes in PTO, coupling strongly to the displacive $\Gamma_4^-$ mode. The phase transition energies are quoted in Table \ref{tab:phasetransitionenergies} and the linear evolution of mode energetics are shown in Figure \ref{fig:ModeEnergy}. \par

\begin{table}[]
\centering
\begin{tabular}{@{}ccccccc@{}}
                               \toprule \toprule                              &                 &                    & SZP   & DZDP  & TZTP  & PW    \\ \midrule
                                                             & Pb 1b           & $T_{1u}$           & 0.000 & 0.000 & 0.000 & 0.000 \\
$\Gamma_4^-$ $\rightarrow$ $\mathbf{q}_{\Gamma} = [0, 0, 0]$ & Ti 1a           & $T_{1u}$           & 0.160 & 0.118 & 0.111 & 0.141 \\
                                                             & O 3d            & $A_{2u}$           & 0.659 & 0.458 & 0.370 & 0.446 \\
                                                             & O 3d            & $E_u$              & 0.842 & 0.646 & 0.550 & 0.651 \\ \midrule
                                                             & \multicolumn{2}{c}{Total distortion} & 1.081 & 0.800 & 0.672 & 0.802 \\ \bottomrule \bottomrule
\end{tabular}
\caption{The mode amplitudes normalised to the parent cell $A_p$ (described in the text) for the irreps characterising the $Pm\bar{3}m \rightarrow P4mm$ phase transition in PTO.}
\label{tab:PTODistortTab}
\end{table}

Before discussing mode amplitudes, we must first carefully define them. We do so following the format of the \texttt{ISODISTORT} $A_p$ amplitude normalised to the primitive cell \cite{Campbell2006}. Once an atomic displacement has been identified as belonging to a particular irrep, the displacement is calculated in fractional coordinates relative to the \textit{parent} structure. This defines the amplitude of a specific displacement in the irrep. To calculate $A_p$ we now normalise the amplitude by a factor of $\sqrt{V_p/V_s}$ for supercell/primitive cell volumes $V_{p/s}$. Now to calculate the amplitude of the irrep as a whole, we take the square root of the sum of the squares of the displacement amplitudes belonging to the irrep in question (thereby obtaining an RMS amplitude). If we wish to characterise the amplitude of the total distortion from the transition, we can take the square root of the sum of the squares for each irrep amplitude. Tables \ref{tab:PTODistortTab} and \ref{tab:PZODistortTab} are then tabulations of $A_p$.

\begin{table}[]
\centering
\begin{tabular}{@{}ccccccc@{}}
                                 \toprule \toprule                          &                      &                          & SZP                       & DZDP                      & TZTP                      & PW                        \\ \midrule
R$_4^+ \rightarrow \mathbf{q}_R = [1/2, 1/2, 1/2]$         & O 3d                 & $E_u$                    & 0.571                     & 0.554                     & 0.553                     & 0.534                     \\
                                                           & \multicolumn{2}{c}{Total R$_4^+$ distortion}    & 0.571                     & 0.554                     & 0.553                     & 0.534                     \\ \midrule
                                                           & Pb 1b                & $T_{1u}$                 & 0.213                     & 0.214                     & 0.239                     & 0.256                     \\
                                                           & Zr 1a                & $T_{1u}$                 & 0.041                     & 0.034                     & 0.037                     & 0.047                     \\
$\Sigma_2 \rightarrow \mathbf{q}_{\Sigma} = [1/4, 1/4, 0]$ & O 3d                 & $A_{2u}$                 & -0.047                    & -0.042                    & -0.035                    & -0.035                    \\
                                                           & O 3d                 & $E_{u_{1}}$              & 0.253                     & 0.250                     & 0.224                     & 0.227                     \\
                                                           & O 3d                 & $E_{u_{2}}$              & -0.230                    & -0.222                    & -0.216                    & -0.217                    \\
\multicolumn{1}{l}{}                                       & \multicolumn{2}{l}{Total $\Sigma_2$ distortion} & \multicolumn{1}{l}{0.407} & \multicolumn{1}{l}{0.400} & \multicolumn{1}{l}{0.396} & \multicolumn{1}{l}{0.409} \\ \midrule
                                                           & Pb 1b                & $T_{1u}$                 & -0.049                    & -0.033                    & -0.026                    & -0.026                    \\
S$_4 \rightarrow \mathbf{q}_S = [1/4, 1/2, 1/4]$           & O 3d                 & $E_{u_{1}}$              & -0.013                    & -0.129                    & -0.103                    & -0.106                    \\
                                                           & O 3d                 & $E_{u_{2}}$              & 0.074                     & -0.083                    & -0.071                    & -0.070                    \\
                                                           & \multicolumn{2}{c}{Total S$_4$ distortion}      & 0.154                     & 0.157                     & 0.128                     & 0.129                     \\ \midrule
                                                           & Pb 1b                & $T_{1u}$                 & -0.006                    & -0.001                    & -0.009                    & -0.010                     \\
M$_5^- \rightarrow \mathbf{q}_M = [1/2, 1/2, 0]$           & Zr 1a                & $T_{1u}$                 & -0.008                    & -0.003                    & -0.006                    & -0.008                    \\
                                                           & O 3d                 & $E_u$                    & 0.012                     & 0.013                     & 0.009                     & 0.009                     \\
                                                           & \multicolumn{2}{c}{Total M$_5^-$ distortion}    & 0.015                     & 0.013                     & 0.014                     & 0.016                     \\ \midrule
                                                           & Pb 1b                & $T_{1u}$                 & 0.011                     & 0.024                     & 0.027                     & 0.027                     \\
R$_5^+ \rightarrow \mathbf{q}_R = [1/2, 1/2, 1/2]$         & O 3d                 & $E_u$                    & -0.001                    & -0.006                    & -0.007                    & 0.008                     \\
                                                           & \multicolumn{2}{c}{Total R$_5^+$ distortion}    & 0.011                     & 0.025                     & 0.028                     & 0.028                     \\ \midrule
                                                           & Zr 1a                & $T_{1u}$                 & 0.001                     & 0.0003                    & 0.002                     & 0.003                     \\
X$_3^- \rightarrow \mathbf{q}_X = [0, 1/2, 0]$             & O 3d                 & $E_u$                    & -0.009                    & -0.009                    & -0.017                    & -0.017                    \\
                                                           & \multicolumn{2}{c}{Total X$_3^-$ distortion}    & 0.009                     & 0.009                     & 0.018                     & 0.017                     \\ \midrule
                                                           & \multicolumn{2}{c}{Overall distortion}          & 0.718                     & 0.702                     & 0.693                     & 0.686  \\   \bottomrule \bottomrule               
\end{tabular}
\caption{The mode amplitudes normalised to the parent cell $A_p$ (described in the text) for the irreps characterising the $Pm\bar{3}m \rightarrow Pbam$ phase transition in PZO. The modes are listed in descending total distortion.}
\label{tab:PZODistortTab}
\end{table}

We consider first the simpler phase transition of PTO. This is characterised by a single displacive mode $\Gamma_4^-$ resulting in the ferroelectric $P4mm$ phase. Examining Table \ref{tab:PTODistortTab} we see that Pb 1b displacements are set to zero since we have chosen the Pb site as the origin. We see that for the SZP basis set the $\Gamma_4^-$ distortion is rather approximate compared to plane-waves overestimated by $\approx 35\%$. This can be better understood if we consider the amplitudes of the strain modes $\Gamma_{1\sigma}^+$ and $\Gamma_{3\sigma}^+$ (the subscript $\sigma$ denotes a strain mode rather than a displacive one). The former is responsible for uniform isotropic expansions/contractions of the cell whilst the latter is responsible for tetragonality. Rather than quote the amplitude of each mode for each basis (which is available in the supplementary material), it is more illuminating to examine the zero pressure lattice constants of Table \ref{tab:latticeconstants}. We see that for the SZP basis the $c/a$ ratio is considerably overestimated at 1.24 when compared to the plane-wave $c/a$ of 1.084. This implies also a considerable overestimate of the amplitudes of $\Gamma_{1\sigma}^+$ and $\Gamma_{3\sigma}^+$. The amplitude of the ferroelectric $\Gamma_4^-$ distortion is then increased as a result of the strong coupling to $\Gamma_{1\sigma}^+$ and $\Gamma_{3\sigma}^+$. A peculiarity of this transition for the default PAOs is their non-systematic nature. Despite the smaller number of basis functions, the DZDP basis performs better than TZTP for the amplitude of displacive modes, strain modes, the phase transition energy (Table \ref{tab:phasetransitionenergies}) and lattice constants of the $P4mm$ phase (Table \ref{tab:latticeconstants}). It is for this reason that great care must be taken when using PAOs to describe systems where the internal and cell degrees of freedom are strongly coupled. We note that using a basis other than the default can result in vast improvements for this transition. When tuning the confinement energies to fit the plane-wave c/a, we can achieve a phase transition energy within 1 meV of the plane-wave energy (shown in the supplemental material). \par

\begin{table}[]
\centering
\scalebox{0.76}{
\begin{tabular}{@{}ccccccccccccc@{}} \toprule \toprule
                                         & \multicolumn{3}{c}{SZP {[}\AA{]}} & \multicolumn{3}{c}{DZDP {[}\AA{]}} & \multicolumn{3}{c}{TZTP {[}\AA{]}} & \multicolumn{3}{c}{PW {[}\AA{]}} \\ \midrule
                                         & a              & b               & c             & a              & b               & c              & a              & b               & c              & a              & b              & c     \\ \midrule
$Pm\bar{3}m$ PTO                   & 3.952          & 3.952           & 3.952         & 3.928          & 3.928           & 3.928          & 3.926          & 3.926           & 3.926          & 3.918          & 3.918          & 3.918         \\
$Pm\bar{3}m$ PZO                   & 4.181          & 4.181           & 4.181         & 4.159          & 4.159           & 4.159          & 4.154          & 4.154           & 4.154          & 4.140          & 4.140          & 4.140         \\
$P4mm$ PTO                         & 3.799          & 3.799           & 4.710         & 3.865          & 3.865           & 4.282          & 3.890          & 3.890           & 4.128          & 3.870          & 3.870          & 4.194         \\
$Pbam$ PZO                         & 5.919          & 11.907          & 8.241         & 5.877          & 11.835          & 8.223          & 5.886          & 11.793          & 8.206          & 5.871          & 11.765         & 8.178         \\
$Fm\bar{3}m$ PZT 50/50 & 8.124          & 8.124           & 8.124         & 8.091          & 8.091           & 8.091          & 8.080          & 8.080           & 8.080          & 8.050          & 8.050          & 8.050         \\
$Pm\bar{3}m$ PZT 50/50 & 8.143          & 8.143           & 8.143         & 8.112          & 8.112           & 8.112          & 8.096          & 8.096           & 8.096          & 8.069          & 8.069          & 8.069         \\ \bottomrule \bottomrule
\end{tabular}}
\caption{The optimised, mutually orthogonal, pseudocubic lattice constants a, b and c for each of the considered crystals and basis set.}
\label{tab:latticeconstants}
\end{table}

The PZO transition is more difficult to unpack. Despite this, (perhaps due to only a weak coupling between displacive and strain modes) the material is impressively described by the default PAOs. The error in both the $Pm\bar{3}m$ and $Pbam$ lattice constants are smaller than 0.5\% for the TZTP basis demonstrating that even highly distorted perovskites can be represented well by the default PAOs. We now examine Table \ref{tab:PZODistortTab} commenting on the individual contributions of each irrep in the transition. Beginning first with the R$_4^+$ mode we see that the accuracy improves with basis set size but accounts for the largest source of error in the overall distortion. Since the R$_4^+$ mode is AFD, this overestimation leads to a larger rotation angle of the oxygen octahedra compared to plane-waves. For the antipolar $\Sigma_2$ distortion, we see that the total distortion becomes \textit{further} from the plane-wave value as we increase the number of basis functions. This is somewhat misleading however as when we examine the individual displacements, we see that O and Pb displacements are better described by the TZTP basis. The reason the SZP basis \textit{appears} to perform better is because the underestimated Pb displacement balances with the O $E_{u_2}$ displacement. Taking both the S$_4$ and X$_3^-$ modes together, we see that an accurate description requires the TZTP basis. It should be noted, however, the X$_3^-$ mode contributes very little to the energetics of the transition which is dominated by the R$_4^+$, $\Sigma_2$ and S$_4$ modes. We note finally that although the amplitudes of some modes does not improve with basis set size, the overall distortion does. By TZTP, the distortion is within 0.1\% of the plane-wave distortion. This accuracy is seen also in the phase transition energy (Table \ref{tab:phasetransitionenergies}) which improves with basis set size within 1\% of the plane-wave value by TZTP.

\begin{table}[]
\centering
\begin{tabular}{@{}lcc@{}}
\toprule \toprule
     & PTO $Pm\bar{3}m$ $\Rightarrow$ $P4mm$ & PZO $Pm\bar{3}m$ $\Rightarrow$ $Pbam$ \\
     & {[}meV/ABO$_3$ unit{]}                & {[}meV/ABO$_3$ unit{]}                \\ \midrule
SZP  & -115.51                               & -295.78                               \\
DZDP & -58.40                                & -256.27                               \\
TZTP & -47.37                                & -258.71                               \\
PW   & -69.83                                & -262.24                               \\ \bottomrule \bottomrule
\end{tabular}
\caption{The phase transition energies for the cubic to tetragonal/orthorhombic transitions in PTO/PZO. This quantity is defined by the difference in total energy between the relaxed cubic phase and the relaxed distorted phase for each basis.}
\label{tab:phasetransitionenergies}
\end{table}

Figure \ref{fig:ModeEnergy} describes the energy difference $\Delta E$ between an initial, undistorted phase and a phase distorted by some fraction of a displacive or strain mode. For each basis, the maximum amplitude of an irrep is the amplitude found in the relaxed structure in Tables \ref{tab:PTODistortTab} and \ref{tab:PZODistortTab}. In Figure \ref{fig:ModeEnergy}a, the initial undistorted structure is the optimised $Pm\bar{3}m$ PTO structure for the basis in question. Since the displacive mode is coupled strongly to the strain modes, we linearly evolve the three modes ($\Gamma_4^-$, $\Gamma_{1\sigma}^+$ and $\Gamma_{3\sigma}^+$) simultaneously until their maximum amplitudes are reached. As was explained previously, the default DZDP basis best approximates the plane-wave energetics since the TZTP basis retrieves a less accurate c/a ratio. A small energy barrier exists for the transition for \textit{all} cases. This is not true in reality, only existing here since we have assumed the displacive and strain modes are directly proportional on a 1:1 basis (the real coupling is non-linear). Notable however, is the size of the barrier for the SZP basis of $\approx 35$ meV. This is an artefact of the large strain modes and the highly non-linear coupling with the displacive mode. For curves along the plane-wave optimised trajectory (explained in the Figure \ref{fig:ModeEnergy} caption), a minimum of energy is always reached before the mode maximum inferring that the optimised plane-wave structure does not so well approximate the optimised structure of a given PAO basis. This, much like the phase transition energy, can be improved upon using a non-default basis. \par

In Figure \ref{fig:ModeEnergy}b, our initial undistorted structure is the optimised orthorhombic $Pbam$ cell of PZO (with lattice constants described in Table \ref{tab:latticeconstants}) but with \textit{zero} displacive mode amplitude. As a general trend, we see that there is a fair discrepancy between the SZP and plane-wave curves. This almost disappears as we use a DZDP basis and even more so by TZTP. A clear source of error in our description for all bases is the overestimated amplitude the R$_4^+$ mode. We see that along the (0, 0, 0) $\rightarrow$ (R$_4^+$, 0, 0) path, this overestimation result in \textit{too much} energy lowering. Interestingly, for the DZDP and TZTP bases, this error seems to recover along the (R$_4^+$, 0, 0) $\rightarrow$ (R$_4^+$, $\Sigma_2$, 0) path and by the time all three displacive modes are present, the energetics are almost identical. For the plane-wave optimised trajectories, we once again observe a minimum of energy before the maximal mode amplitude. This barrier appears just before the (R$_4^+$, $\Sigma_2$, 0) point indicating that the plane-wave $\Sigma_2$ mode (when coupled to R$_4^+$ at least) does not describe a stable structure. This effect does appear to diminish with basis set size, disappearing almost entirely for TZTP. At the (final) (R$_4^+$, $\Sigma_2$, S$_4$) point, we see that the DZDP, DZDP(PW), TZTP and TZTP(PW) curves very well represent the PW curve. This implies that in this case, the default PAOs could be used to describe a plane-wave optimised structure with only a very small penalty in the energetics. This is valuable since this allows us take small plane-wave optimised cells to build larger supercells for PAO calculations \textit{without re-relaxing} the structure allowing for easy up-scaling of accurate DFT calculations. 
    
\section{\label{Summary:level1}{Summary}}

We have investigated the consequences of representing delicate features of the perovskite oxides with the default \textsc{CONQUEST} basis sets of PAOs as a replacement for plane-waves. In the electronic structure, we were able to reproduce the plane-wave electronic charge density with an error of $\approx 0.5$\% using the total integrated electronic error integral of equation \ref{eq:NeError}. We found that the largest source of error to this integral is from a surplus of electronic density close to O anion sites as shown in electronic charge density difference plots. Even fine features derived from Bader partitioning (Bader charges, volumes and densities) agree well with plane-wave calculations and once again demonstrates the small surplus in electronic density near O anion sites. We quantified the completeness of the PAO basis sets by providing plane-wave cutoff energies offering the same accuracy in $N_{\text{error}}$ and energy convergence as those using the plane-wave basis. We found that although the two metrics disagree (by a small amount) on the cutoff, by TZTP we can achieve the accuracy of a 27.28-30.85 Ha plane-wave cutoff, close to \textit{double} what has previously been reported in the literature. We note that whilst this comparison is useful, the error cancellation in the plane-wave basis (that is, error in the core regions tend to cancel with one another) is not perfectly achieved when we consider the energy difference between plane-wave and PAO calculations. Further, we do not expect these errors to be system dependent in the case of plane-waves but could be for PAOs, especially smaller basis sets (like SZ and SZP). \par

When investigating the condensation of soft-modes, we found that larger basis sets of PAOs (DZDP and TZTP) well described the $Pm\bar{3}m \rightarrow Pbam$ phase transition in PZO, both structurally and energetically. Impressively, both the total distortion and phase transition energy when using the TZTP basis is within 1\% of the plane-wave figures. We found that more care had to be taken for the $Pm\bar{3}m \rightarrow P4mm$ phase transition in PTO due to the strong coupling between displacive and strain modes. This, however, can be remedied by using a non-default optimised basis. Remarkably, we find that using a default DZDP or TZTP basis set on a plane-wave optimised geometry for PZO results in close to identical phase transition energies. This suggests that even highly distorted perovskites can be represented by basis sets of default PAOs. \par

This work suggests that even fine structural and electronic features of the perovskite oxides can be calculated to near plane-wave accuracy using basis sets of PAOs. Since PAO-based calculations (in partnership with the correct algorithm) can scale to many thousands (and millions \cite{Bowler2010}) of atoms, this approach now offers a pathway for accurate large scale first principles simulations of perovskite systems using DFT. This is particularly valuable as many research questions (as discussed in Section \ref{sec:intro}) facing the perovskite oxides require many thousands of atoms to investigate. 

\section*{Acknowledgements}
The authors are grateful for computational support from the
UK Materials and Molecular Modelling Hub, which is partially funded by EPSRC (EP/P020194), for which access was
obtained via the UKCP consortium and funded by EPSRC
Grant ref EP/P022561/1. We are also thankful for the computational resources provided by the London Centre for Nanotechnology, including access to the Salviati computing cluster.

\section*{References}
\bibliographystyle{ieeetr}
\bibliography{ms.bib}

\end{document}


\author{Jack S. Baker$^{1, 2}$, Tsuyoshi Miyazaki$^3$ \& David R. Bowler$^{1, 2, 3}$  \\\small{$^1$\textit{London Centre for Nanotechnology, UCL, 17-19 Gordon St, London WC1H 0AH, UK}} \\ \small{$^2$\textit{Department of Physics \& Astronomy, UCL, Gower St, London WC1E 6BT, UK}} \\
\small{$^3$\textit{International Centre for Materials Nanoarchitectonics (MANA)}},\\ \small{\textit{National Institute for Materials Science (NIMS)}},\\ \small{\textit{1-1 Namiki, Tsukuba, Ibaraki 305-0044, Japan}}}

\title{\textbf{Supplemental Material:} \\ \medskip \Large{The pseudoatomic orbital basis: electronic accuracy and soft-mode distortions in ABO$_3$ perovskites}}
\maketitle

\medskip


This document provides all the necessary supplemental material for the article "\textit{The pseudoatomic orbital basis: electronic accuracy and soft-mode distortions in ABO$_3$ perovskites}". It is divided into two sections. Section \ref{sec:basisdetails} supplements the details of the default basis used in the main text with a figure of the radial functions as well as detailing an optimised basis set including a description of the optimisation procedure. Section \ref{sec:strainmodes} includes a tabulation of the relevant strain mode amplitudes in the $Pm\bar{3}m \rightarrow P4mm$ and $Pm\bar{3}m \rightarrow Pbam$ phase transitions of PTO and PZO respectively. If further information or raw data are required, do not hesitate to contact the authors via email\footnote[2]{Jack S. Baker: \textcolor{blue}{jack.baker.16@ucl.ac.uk}, Tsuyoshi Miyazaki: \textcolor{blue}{MIYAZAKI.Tsuyoshi@nims.go.jp} \\ David R. Bowler: \textcolor{blue}{drb@ucl.ac.uk}}.

\section{\label{sec:basisdetails} Basis sets: defaults and optimised}

Figure \ref{fig:RadialFuncs} shows the radial functions for the default PBESol \cite{Perdew2008} basis used in this work. All basis sets are generated using the \textsc{CONQUEST} PAO generator code (\texttt{v1.02}) \cite{CQRelease2020} using the default settings of the equal radii method of construction as described in the main text. That is, the SZP basis is obtained in the code by setting the \texttt{Atom.BasisSize} flag to \texttt{small} and the TZTP basis is obtained by setting \texttt{Atom.BasisSize} to \texttt{large}. Strictly, the code currently has no flag to generate a DZDP default, but, if we set \texttt{Atom.BasisSize} to \texttt{medium} we get a default DZP basis. Upon including the second zeta from the polarisation functions in the TZTP basis, we arrive at our default DZDP basis. Following Figure \ref{fig:RadialFuncs}, the SZP basis set used in this work is comprised of just $\zeta = 1$ PAOs whilst the DZDP basis includes also the $\zeta = 2$ functions and TZTP adds on the $\zeta = 3$ functions. \par

In the main text, we make reference to the use of an optimised basis performing much better than the default for the $Pm\bar{3}m \rightarrow P4mm$ phase transition energy of PTO. We performed this optimisation using a heuristic approach aiming to minimise the difference in the c/a ratio of $P4mm$ PTO obtained from a DZDP PAO basis with the plane-wave result using the local density approimation of Perdew \& Wang \cite{Perdew1992}. We choose the DZDP basis since this offers good flexibility whilst having few enough zetas (and therefore parameters in the optimisation process) to reliably reach a minumum. In this process, the cutoff radii $r_c$ (or equivalently, the confinement energies) of each PAO are the variables of the minimisation. Fitting to the c/a ratio is a good choice for the ferroelectric phase of PTO since it is indicative of the bulk polarisation. The optimisation procedure can be described with three stages:

\begin{figure}
    \centering
      \includegraphics[width=\linewidth]{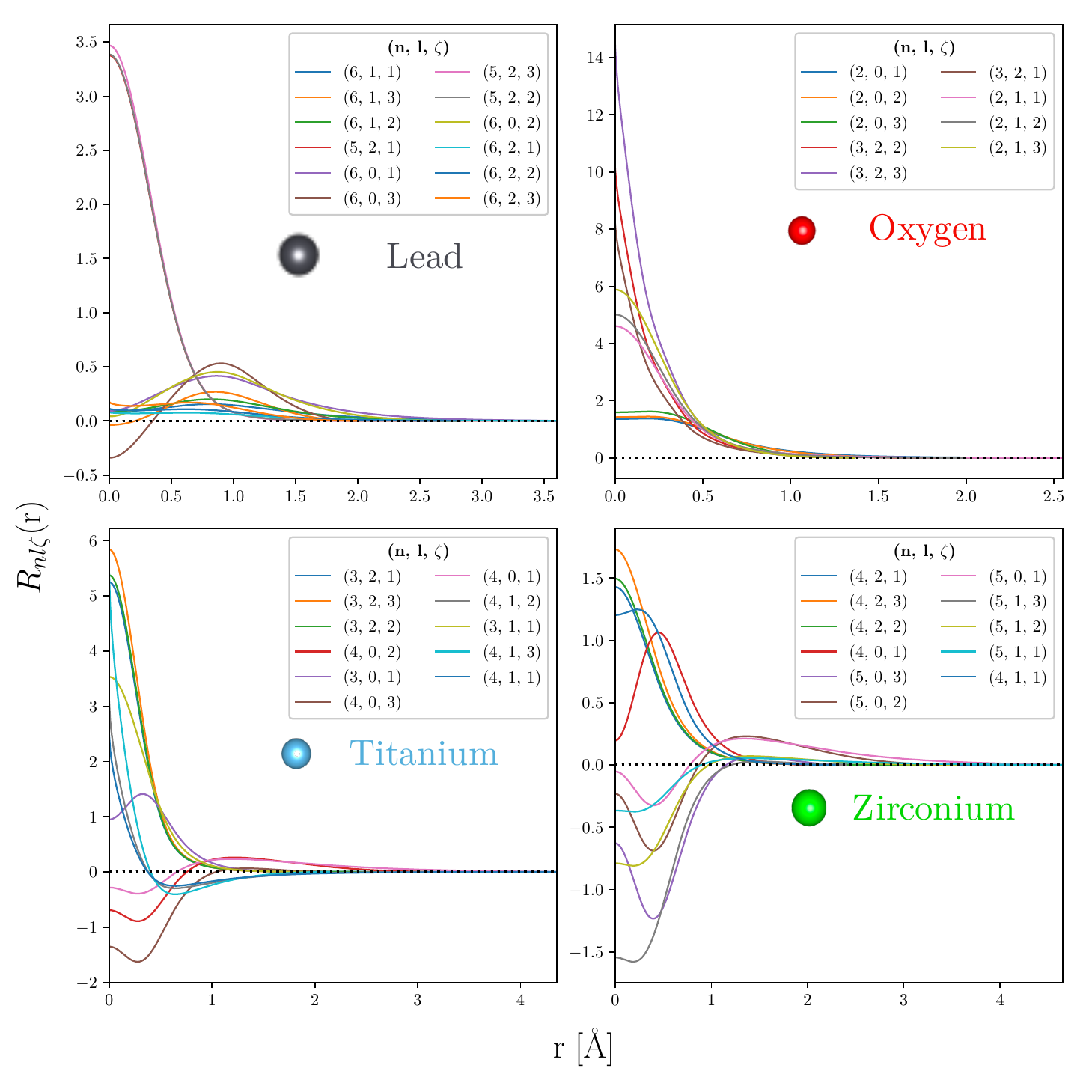}
        \caption{The default radial function solutions $R_{nl\zeta}(r)$ solved in the pseudoatom potential for lead (top left), oxygen (top right) titanium (bottom left) and zirconium (bottom right). We include radial functions for $\zeta$ = 1, 2 \& 3 which when multiplied with the appropriate spherical harmonics create the SZP, DZDP and TZTP default basis sets used in this work. }
        \label{fig:RadialFuncs}
    \end{figure}
\begin{enumerate}

    \item Solve for the radial functions $R_{nl\zeta}(r)$ for an initial (the default) set of $r_c$ for each species for the confined pseudoatom.
    
    \item Using these PAOs, Perform a full geometry optimisation of the atomic positions and cell size/shape starting from the plane-wave optimised $P4mm$ PTO structure.
    
    \item If the resulting c/a ratio agrees with the plane-wave value to some tolerance, terminate. If not, return to stage 1 with a modified set of $r_c$.
    
\end{enumerate}

The resulting cutoff radii $r_c$ for the radial functions of the LDA optimised basis are shown table \ref{tab:optrc}. Rather usefully, the obtained cutoff radii are compact in their range, so, can be used for efficient large scale calculations. The optimised DZDP phase transition energy for this functional (-47.91 meV) is accurate to the plane-wave calculation (-48.12 meV) as calculated with \texttt{ABINIT} (v8.10.2) \cite{Gonze2009,Gonze2016} by +0.44\% whilst the optimised DZDP c/a ratio (1.036) is accurate to -0.19\%. This is an improvement on the default DZDP basis (by the equal radii method) which has a c/a ratio of 1.030 and a phase transition energy of 17.95 meV. Much like the default TZTP PBESol basis used in the main text, the default TZTP LDA basis does \textit{not} improve on the default DZDP basis, underestimated the c/a ratio further. We note that whilst our optimisation method clearly offers vast improvements for this application, our approach is \textit{not} generally reliable for all applications and is prone to encountering local minima as well as interfering with the transferability of the basis. It is also fair to point out however that at the time of publication, these same issues plague all current methods of PAO optimisation.

\begin{table}[]
\centering
\begin{tabular}{@{}ccc@{}}
\toprule \toprule
      & $r_c$, $\zeta=1$ {[}\AA{]} & $r_c$, $\zeta=2$ {[}\AA{]} \\ \midrule
Pb 5d & 2.963                                     & -                                         \\
Pb 6s & 3.498                                     & 2.069                                    \\
Pb 6p & 3.461                                     & 2.148                                     \\
Pb 6d & 2.069                                     & 1.413                                     \\
Ti 3s & 2.397                                     & -                                         \\
Ti 3p & 2.704                                     & -                                         \\
Ti 4s & 3.043                                     & 1.699                                     \\
Ti 3d & 3.006                                     & 1.656                                     \\
Ti 4p & 2.127                                     & 1.598                                     \\
O 2s  & 2.662                                     & 1.434                                     \\
O 2p  & 3.186                                     & 1.773                                     \\
O 3d  & 3.186                                     & 1.773                                     \\ \bottomrule \bottomrule
\end{tabular}
\caption{The resulting cutoff radii $r_c$ for the optimised DZDP basis using the heuristic optimisation process detailed above. Orbitals not detailing a second $\zeta$ have only a single $\zeta$ in the basis.}
\label{tab:optrc}
\end{table}

\section{\label{sec:strainmodes} Strain modes}
\begin{table}[]
\centering
\begin{tabular}{@{}cccccc@{}}
\toprule \toprule
    &                      & \multicolumn{4}{c}{Mode amplitude, $A_p$} \\ \midrule
    & irrep                & SZP      & DZDP     & TZTP     & PW       \\ \midrule
PTO & $\Gamma_{1\sigma}^+$ & 0.0667   & 0.0321   & 0.0166   & 0.0266   \\
    & $\Gamma_{3\sigma}^+$ & 0.1883   & 0.0867   & 0.0495   & 0.0674   \\
PZO & $\Gamma_{1\sigma}^+$ & -0.0037  & -0.0037  & -0.0037  & -0.0029  \\
    & $\Gamma_{3\sigma}^+$ & -0.0150  & -0.0115  & -0.0124  & -0.0132  \\
    & $\Gamma_{5\sigma}^+$ & 0.0041   & 0.0049   & 0.0013   & 0.0014   \\ \bottomrule \bottomrule
\end{tabular}
\label{tab:strainmode}
\caption{The strain mode amplitudes, $A_p$, for the $Pm\bar{3}m \rightarrow P4mm$ phase transtion of PTO and the $Pm\bar{3}m \rightarrow Pbam$ phase transition of PZO}
\end{table}

We considered only displacive modes in the main text for the phase $P4mm$ \& $Pbam$ phase transitions of PTO \& PZO respectively. These modes are also coupled to strain modes whose amplitudes ($A_p$) are tabulated in Table \ref{tab:strainmode} following the definition from \texttt{ISODISTORT} \cite{Campbell2006} (which is also described in the main text). The $P4mm$ PTO transition involves two strain modes, $\Gamma_{1\sigma}^+$ (responsible for isotropic expansion/contraction) and $\Gamma_{3\sigma}^+$ (responsible for a contracting $a$-axis and elongating $c$-axis, causing tetragonality) whilst the $Pbam$ PZO transition involves three strain modes $\Gamma_{1\sigma}^+$ (responsible for isotropic expansion/contraction), $\Gamma_{3\sigma}^+$ (responsible for a contracting $c$-axis and shortening $b$-axis and mildy contracting $a$-axis causing orthorhombicity) and $\Gamma_{5\sigma}^+$ (resposible for a contracting $a$-axis and expanding $b$-axis, causing tetragonality)

We see that the strain modes for PTO tell a similar story to what is said in the main text which uses the c/a ratio as a surrogate in the discussion. The SZP basis features strain modes more than doubling the plane-wave result, leading to a large overestimate in the tetragonality (and the bulk polarisation). The DZDP basis still overestimates these modes whilst TZTP underestimates them. For PZO, we see that the amplitudes are \textit{much} smaller than those participating in the PTO transition (the reason for which they are not is discussed in the main text). It can be seen that both the SZP and DZDP basis sets describe adequately the $\Gamma_{1\sigma}^+$ \& $\Gamma_{3\sigma}^+$ modes whilst a good description of $\Gamma_{5\sigma}^+$ is not reached until TZTP. This is likely of little merit however due to the the tiny amplitude of the mode.

\bibliographystyle{ieeetr}
\bibliography{supplement.bib}